\documentclass[aps,showkeys,groupedaddress,twocolumn,showpacs]{revtex4/revtex4}
\usepackage{graphicx}
\usepackage[ansinew]{inputenc}
\usepackage[tbtags]{amsmath}
\usepackage{amssymb}

\newcommand{\erfc}{\mbox{erfc}}
\newcommand{\joint}{$\rho^{\Theta}(x_{n},x_{n+1},d)$ }

\newcommand{\ma}{$\rho(x_n|X(\eta),a)$ }

\newcommand{\antima}{$\rho(x_n|\overline{X(\eta)},a)$ }
\begin{document}
\title{Precursors of extreme increments}
\author{Sarah Hallerberg, Eduardo G. Altmann, Detlef Holstein, Holger Kantz}  
\affiliation{Max Planck Institute for the Physics of Complex Systems\\
N\"othnitzer Str.\ 38, D 01187 Dresden, Germany\\}
\date{\today}
\begin{abstract}
We investigate precursors and predictability of extreme increments in a time
series. The events we are focusing on consist in large increments within
successive time steps. We are especially interested in understanding how the quality of the predictions depends on the strategy to choose precursors,
on the size of the event and on the correlation strength. We study the
prediction of extreme increments analytically in an AR(1) process, and
numerically in wind speed recordings and long-range correlated ARMA data.
We evaluate the success of predictions via
receiver operator characteristics (ROC-curves). 
Furthermore, we observe an increase of the quality of predictions with
increasing event size and with decreasing correlation in all examples. Both effects can be understood by using the likelihood ratio as a summary index for smooth ROC-curves.
\end{abstract}
\maketitle
\section{Introduction}

Extreme value statistics \cite{Coles} is a well established approach to predict
the relative frequency of rare extreme events, but does not include forecasts
of when the next event will occur.
There have been many attempts to employ time series
strategies for the latter purpose. These strategies usually investigate a record of
historical data about the phenomenon under study and try to infer knowledge
about the future. A standard approach is to search for precursors, i.e.,
typical signatures preceeding an extreme event. Such precursors have
been discussed, e.g., in the literature about earthquakes \cite{Jackson}, 
epileptic seizures  \cite{Kapiris}, and stock market crashes
\cite{Sornette2,stocks,Sornette3}. As the above listed examples illustrate, the definitions of
what an extreme event is depends on the context. Frequently, one encounters
extremely large values of some observable, or some drastic changes. It is the
latter which is the focus of this paper where we discuss large increments
motivated by stock markets or by turbulent gust in wind speed data.

One might expect that the more extreme an event is, the more difficult it is
to predict it, simply because more extreme events are usually also much
rarer.  However, it has been reported in the literature of wind speed
predictions \cite{Physa}, precipitation forecast \cite{Goeber}, multi agent games \cite{Johnson1} and earthquakes
\cite{Fatemeh} that more extreme events
are better predictable than small events. Therefore one particular goal of
this contribution is to investigate how the predictability of large increments
depends on the size of the increment. 
 
In this contribution we study predictions in a simple autoregressive process
of order 1 (AR(1) process) \cite{Box-Jen} analytically in order to obtain a detailed understanding of
some questions on precursors and predictions. The AR(1) process is a simple
stationary stochastic model process, that might not reflect all features of
more complex processes occurring in nature, but it admits a fully analytic
treatment. Additionally, we study similar prediction procedures numerically
in long-range correlated data and in wind speed data, verifying the same
quantitative results.
The questions, which we intend to answer are the following:
\begin{itemize}
\item [{\bf (Q1)}] How to choose a precursor in order to obtain good
predictions? 
\item [{\bf (Q2)}] Are extreme increments the better predictable, the more
extreme they are?
\item [{\bf (Q3)}] How does the correlation of the data influence the predictability
of extreme increments?
\end{itemize}

The paper is organized as follows.
In Sec.\ \ref{pre} we discuss two strategies which can be used to choose
precursory structures and in Sec.\ \ref{roc} we introduce a method to evaluate
the predictive power of precursors.  The extreme events we dicuss in this
contribution are defined in Sec.\ \ref{def} and we show how to obtain their
joint PDFs analytically in Sec.\ \ref{tfilter}. We apply these procedures to 
AR(1) correlated stochastic processes in Sec.\ \ref{AR1}, to wind speed
measurements in Sec.\ \ref{wind} and to long-range correlated data Sec.\ \ref{long}. Conclusions appear in Sec.\ \ref{con}.
\section{Definitions and set-up\label{sec2}}
The considerations in this introductory section are made for general dynamical
systems
with a complex time evolution. They might be purely deterministic, then high-dimensional and
chaotic, or they might be stochastic. In any case we assume that the
time evolution of the system cannot be easily modeled and hence one tries to
extract information about the future from time series data.
 This means that through some experimental observation one can record a usually
univariate time series, i.e., a set of measurements $x_n$ at discrete
times $t_n$, where $t_n = t_0 + n\Delta$ with a sampling
interval $\Delta$. The recording should contain sufficiently many extreme events so that we are able to extract statistical
information about them. We also assume that the event of
interest can be identified on the basis of the observations, e.g., by the
value of the observation function exceeding some threshold, by a sudden
increase, or by its variance exceeding again some threshold.
\subsection{The choice of the precursor \label{pre}}
Ideally, a precursor is a typical signature in the data preceeding {\em every}
individual event. Unfortunately the time evolution of
most systems is usually too irregular to demand this, so one would call
a precursor a data structure which is {\em typically} preceeding an event,
allowing deviations from the given structure, but also allowing events without
preceeding structure. This interpretation of a precursor allows to determine
the specific values of the precursory structure by statistical considerations. 

In order to predict an event occurring at the time $(n+1)$ we compare the last
$k$ observations ${\mathbf x}_{(n,k)}= (x_{n-k+1},x_{n-k+2}, ..., x_{n-1},x_n)$
with a specific precursory structure ${\mathbf x}_{pre}= (x_{n-k+1}^{pre},
x_{n-k+2}^{pre}, ...,x_{n-1}^{pre}, x_n^{pre})$.

This precursory structure can be chosen according to different strategies. The
two possible strategies which we address here, represent the
most fundamental choices. They consist in using either the
maximum of the {\sl a posteriori PDF} or of the maximum of the {\sl
likelihood}  \cite{Bernado}.
In more applied examples one looks for precursors which minimize or maximize
more sophisticated quantities, e.g., discriminant functions or loss matrices.
These quantities are usually functions of the posterior PDF or the
likelihood, but they take into account the additional demands of the
specific problem, e.g., minimizing the loss due to a false prediction \cite{Bishop}.
The two strategies studied in this contribution are thus fundamental in the
sense that they enter into most of the more sophisticated quantities which
are used for predictions and decision making.

The a posteriori PDF $\rho({\mathbf x}_{(n,k)}|X)$ takes
into account all events of size $X$ and provides the
probability density to find a specific precursory structure before an observed
event. 

\begin{itemize}
\item [(I)]  Hence strategy I consists in defining the precursors in a
retrospective or {\sl a posteriori} way: once the extreme event $X$ has been
identified, one asks for the signals right before it.
Formally, this implies that the precursory structure consists of the global maxima in
 each component $(x_{n-k+1}^{*}, x_{n-k+2}^{*},...,x_{n-1}^{*}, x_n^{*})$ of
the a posteriori PDF.
\end{itemize}

The likelihood  $\rho(X|{\mathbf x}_{(n,k)})$ takes into account all possible
values of precursory structures, and provides the probability density that an
event of size X will follow them. Note that the likelihood is thus not a
density function with respect to the precursory structure, but with respect to
the event size X. The precursory structure enters into the likelihood only as a
parameter. 

\begin{itemize}
\item [(II)] Strategy II consists in determining those values of each
component $x_i$ of the condition ${\mathbf x}_{(n,k)}$ for which the
likelihood has a global maximum.
\end{itemize}
Note that the a posterior PDF and the likelihood are linked via Bayes's
theorem
\begin{eqnarray}
 \rho({\mathbf x}_{(n,k)}, X) & = & \rho({\mathbf x}_{(n,k)})\, \rho(X|{\mathbf x}_{(n,k)}) =
\rho({\mathbf x}_{(n,k)}|X)\,\rho(X), \nonumber 
\end{eqnarray}
where $\rho({\mathbf x}_{(n,k)})$ represents the marginal PDF to find the
precursory structure ${\mathbf x}_{(n,k)}$  and $\rho(X)$ represents the marginal
PDF to find events of size X.

In summary the possible values of precursors are given by
\begin{eqnarray}
{\mathbf x}_{pre} & = & \left\{ \begin{array}{l} {{\mathbf x}_{I}}, \\
{{\mathbf x}_{II}},
 \end{array}\label{precursor} \right.\\
\mbox{where}\quad
{\mathbf x}_{I} & := &  \left( x_{n-k+1}^{*},
x_{n-k+2}^{*},...,x_{n-1}^{*},x_n^{*}\right), \nonumber\\
  \mbox{and} \quad {\mathbf x}_{II} & := &  \left(x_{n-k+1}^{\dagger}, x_{n-k+2}^{\dagger},...,x_{n-1}^{\dagger},x_n^{\dagger}\right),\nonumber
\end{eqnarray}
where $x_i^*$ are the points in which $\rho({\mathbf x}_{(n,k)}|X)$ has a global maximum
and $ x_i^{\dagger}$  are the points in which $\rho(X|{\mathbf x}_{(n,k)})$ has
its largest
 maximum, with $n-k+1\leq i\leq n$. In both cases the event size $X$ is assumed to be fixed.
Once the precursory structure  ${\mathbf x}_{pre}$ is determined, we give an alarm for an extreme event when we find the last $k$ observations
${\mathbf x}_{(n,k)}$ in the volume 
\begin{widetext}
\begin{eqnarray}
V_{pre}(\delta) & = &
\left(x_{n-k+1}^{pre}-\frac{\delta}{2},x_{n-k+1}^{pre}+\frac{\delta}{2}\right)\times \left(
x_{n-k+2}^{pre}- \frac{\delta}{2} , x_{n-k+2}^{pre}+ \frac{\delta}{2} \right) \times
... \times \left(x_n^{pre}-\frac{\delta}{2},x_n^{pre}+\frac{\delta}{2}\right). \label{vol}
\end{eqnarray}
\end{widetext}
This method of determining the precursor is especially useful if the PDF of
a process has one clearly defined maximum. For multimodal PDFs the strategy of
using only the global maxima can surely be
improved by considering also the influence of smaller maxima of the
PDF. In this case the precursory volume could, e.g., consist of ${\mathbf x}_{(n,k)}$ for which the PDFs have
values above a certain threshold. In this case $V_{pre}(\delta)$ might not be
simple connected, but apart from this the procedure of predicting should not
be different. However, we restrict ourselves to unimodal PDFs in this contribution.
\subsection{Testing for predictive power\label{roc}}
A common method to verify a hypothesis or test the quality of a prediction is
the receiver operating characteristic curve (ROC-plot) \cite{Swets1,Egan}.
The idea of the ROC-curve consists simply in comparing the rate of correctly
predicted events $r_{c}$ with the rate of false alarms $r_{f}$ by plotting
$r_c$ vs. $r_f$. The resulting curve in the unit-square of the $r_f$-$r_c$
plane  approaches the origin for $\delta \rightarrow0$ and the point $(1,1)$ in the limit $\delta
\rightarrow \infty$, where $\delta$ accounts for the size of the precursor
volume $V_{pre}(\delta)$ (see Eq.\ (\ref{vol})). 

The shape of the curve characterizes the significance of the prediction. A
curve above the diagonal reveals that the corresponding strategy of prediction
is better than a random prediction which is characterized by the
diagonal. Furthermore we are interested in curves which converge as fast as
possible to $r_c=1$, since this scenario tells us that we reach the highest
possible rate of correct prediction without having a large rate of false
alarms.

 There are various so called  {\it summary indices} \cite{Pepe}
which quantify the behavior of the ROC. 
%
%
In this contribution we use the so called {\it likelihood ratio} \cite{Egan} in
order to quantify the ROC-curve. The likelihood ratio
is identical to the slope $m$ of the ROC-curve. For the usage as a summary index,
we consider the slope in the vicinity of the
origin which implies $\delta \rightarrow 0 $.

The term likelihood ratio results from signal detection theory in which context
 the term "a posteriori PDF" refers to the PDF
which we call likelihood in the context of predictions, and vice versa. This is
due to the fact that the aim of signal detection is to identify a signal
which was already observed in the past, whereas predictions are made about
future events. Thus the
"likelihood ratio" is in our case in fact a ratio of the posterior PDFs, as
defined by
\begin{eqnarray}
m & = & \frac{\Delta r_c}{\Delta r_f} \sim \left. \frac{\rho({\mathbf x}_{(n,k)}|X)}{\rho({\mathbf x}_{(n,k)}|\overline{X})}
\right|_{\delta \approx 0} + \mathcal{O}(\delta)\label{defm},
\end{eqnarray}
where $\rho({\mathbf x}_{(n,k)}|\overline{X})$ denotes the a posterior PDF for non-events.
However, we will use the common name likelihood ratio throughout the text.

The {likelihood ratio} can be
expressed in terms of the likelihood $\rho\bigl(X|{\mathbf x}_{(n,k)}\bigr)$ and the total probability to find events $\rho\bigl(X\bigr)$
\begin{eqnarray}
m({\mathbf x}_{(n,k)},X) & \sim & \frac{\Bigr(1-\rho(X)\Bigr)}{\rho(X)}
\frac{\rho(X|{\mathbf x}_{(n,k)})}{\Bigl(1 - \rho(X|{\mathbf x}_{(n,k)})\Bigr)}. \label{mint}
\end{eqnarray}
If we assume that the events we are observing are quite rare and hence $\rho(X),
\rho(X|{\mathbf x}_{(n,k)}) \ll 1$, the likelihood ratio is approximately given by

\begin{eqnarray}
m ({\mathbf x}_{(n,k)},X)&\sim& \frac{\rho(X|{\mathbf x}_{(n,k)})}{\rho(X)} =
\frac{\rho({\mathbf x}_{(n,k)}|X)}{\rho({\mathbf x}_{(n,k)})} 
\quad \label{shortm}
\end{eqnarray}
Eq.\ \ref{shortm} already suggest an answers to questions {\bf (Q1)} and
{\bf (Q2)}, by considering $m ({\mathbf x}_{(n,k)},X)$ as a summary index.
{\bf ad (Q1):} This asymptotic form of the likelihood ratio allows us to
compare different strategies of prediction. Looking for the maximum
of $\rho({\mathbf x}_{(n,k)}|X)$ in ${\mathbf x}_{(n,k)}$, according to
strategy I, there is always the influence of the denominator
$\rho({\mathbf x}_{(n,k)})$ which will keep the likelihood ratio small, even if
$\rho({\mathbf x}_{(n,k)}|X)$ in ${\mathbf x}_{(n,k)}$ is maximized. This is
due to the fact that $\rho({\mathbf x}_{(n,k)}|X)$ cannot be large without
$\rho({\mathbf x}_{(n,k)})$ being large. Strategy II, which uses
the maximum of $\rho(X|{\mathbf x}_{(n,k)})$ in $ {\mathbf x}_{(n,k)} $ should thus be superior, since
the denominator $\rho(X)$ is independent of the chosen precursor. The examples
which are studied in Sec.\ \ref{AR1}, Sec.\ \ref{wind} and Sec.\ \ref{long} support this idea.

{\bf ad (Q2):}
According to Eq.\ (\ref{shortm}), the likelihood ratio is larger than unity, if
$\rho({\mathbf x}_{(n,k)},X) > \rho({\mathbf x}_{(n,k)})\rho(X) $, i.e., if $ {\mathbf x}_{(n,k)}$
and $X$ are correlated. This condition can be also written as $
\rho(X|{\mathbf x}_{(n,k)}) > \rho(X)$ or as $\rho({\mathbf x}_{(n,k)}|X) >
\rho({\mathbf x}_{(n,k)})$ using Bayes's theorem. The latter expression states
that the a posteriori PDF $\rho({\mathbf x}_{(n,k)}|X)$, i.e., the probability to
find the precursor prior to an event should be larger than the probability to
find the precursor prior to an arbitrary value.
Thus, the condition is
fulfilled by choosing the precursor in a reasonable way, e.g.,
using the maximum of $\rho({\mathbf x}_{(n,k)}|X)$  in ${\mathbf x}_{(n,k)}$ or the
maximum of $\rho({\mathbf x}_{(n,k)}|X)$.

\subsection{Definition of Extreme Increments \label{def}}
 In this contribution we will concentrate on extreme events which
consist in a sudden increase (or decrease) of the observed variable within a
few time steps. Examples of this kind of extreme events are the increases in
wind speed in \cite{Physa,Euromech}, but also stock market
crashes \cite{stocks,Sornette2} which consist in sudden decreases.

We define our extreme event by an increment $x_{n+1}-x_n$
exceeding a given threshold $d$
\begin{equation}
 x_{n+1}-x_n \geq   d, \label{e0}
\end{equation}
where $x_{n}$ and $x_{n+1}$ denote the observed values at two consecutive time
steps.

\subsection{Obtaining the analytic expression of the posterior PDFs \label{tfilter}}

A mathematical expression for a filter, which selects the PDF of our
extreme events out of the PDFs of the underlying stochastic process can be
obtained through the  Heaviside function $ \Theta( x_{n+1} - x_{n}
-d)$. This filter is then applied to the joint PDF of a stochastic process.

Since only the time steps $(x_n, x_{n+1})$ are of relevance for the filtering,
we can neglect all previous time steps and apply the filter simply to the
joint PDF for $(x_n, x_{n+1})$, which has then the form $\rho(x_n, x_{n+1}) =
\rho(x_n) \rho(x_{n+1}|x_n)$ 
 This implies that we can regard all previous time-steps $x_0, x_1, ...,
x_{n-1}$, on which $\rho_n$ and $\rho_{n+1}$ might depend, as
parameters.

The joint PDF of the extreme events $\rho^\Theta(x_{n+1},x_n,d)$ can then be
obtained by multiplication with $\Theta( x_{n+1} - x_{n}-d)$. If the resulting
expression is non zero, the condition of the extreme event (\ref{e0}) is
fulfilled and for $x_{n+1}$ and $x_{n}$ the following relation holds:
\begin{eqnarray}
x_{n+1} & = & x_{n} + d + \gamma \label{gammadef} \quad (\gamma \in  \mathbb R,
\gamma \geq 0) \quad. \label{gamma0}
\end{eqnarray}
Hence it is possible to express the joint probability density in terms of
$x_{n}$ or $x_{n+1}$ with the new random variable $\gamma$. We can then use the integral representation of the Heaviside function with appropriate
substitutions to obtain: 
\begin{eqnarray}
f^\Theta(x_{n+1},x_{n},d)&=& \rho(x_{n})\int_{0}^{\infty} \rho(x_{n} + d + \gamma|x_n)
\nonumber\\
&\it{}&\quad \delta((x_{n+1} - x_{n}
- d) - \gamma)\;d\gamma.\quad\label{int1} 
\end{eqnarray}
By normalizing this expression with the total probability $\rho_{\Theta} (d)$ to find extreme
events of size $d$ or larger
we obtain the joint PDF \joint of all values of $x_{n}$ and $x_{n+1}$ which are part of an extreme event.
Integrating the resulting joint PDF \joint over $x_{n+1}$ we find the
following expression for the marginal distribution, i.e., the a posteriori PDF: 
\begin{eqnarray}
 \rho(x_n|X(d))
 & = &  
\frac{\rho(x_{n})}{\rho_{\Theta}(d)}\int_{0}^{\infty}d\gamma\;\rho(x_{n}
+ d + \gamma|x_n). \nonumber\\ \label{marginal}
\end{eqnarray}

Analogously $\rho(x_{n}|\overline{X(d)})$ denotes the a posteriori PDF to
observe the value $x_{n}$ before an non-event, i.e., before an increment which
is smaller than $d$.

\begin{eqnarray}
\rho(x_{n}|\overline{X(d)}) & = &
\frac{\rho(x_{n})}{(1-\rho_{\Theta}(d))}\int_{-\infty}^{\infty} dx_{n+1}\; \Bigr(1- \Bigl. \nonumber\\
&\it{}& \;\;\;\;\Theta(x_{n+1}-x_{n}-d)\Bigl) \rho_{n+1}(x_{n+1}|x_{n}). \nonumber \\ 
\end{eqnarray}

If for a given process the joint PDF of two consecutive events is known, we can hence analytically
determine $\rho(x_n|X(d))$, $\rho(x_{n}|\overline{X(d)})$ and
$\rho_{\Theta}(d)$.

\section{Extreme increments in the AR(1) model \label{AR1}}
\subsection{AR(1) model}
\begin{figure}[t!!!]
\includegraphics[width=6cm, angle= -90]{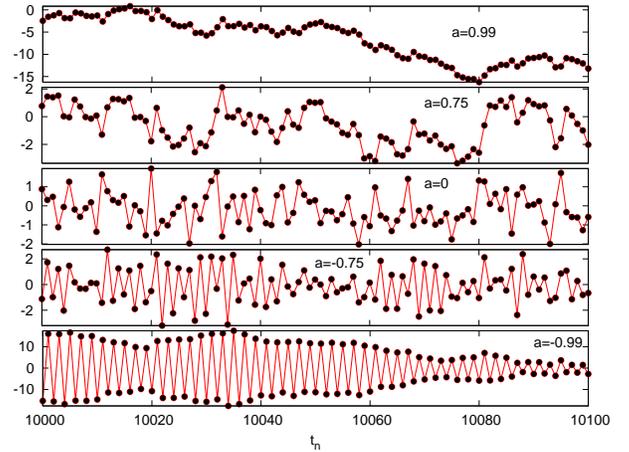} 
\caption[]{\small \label{fig:data+} (Color online)
Parts of the time series of the AR(1) process for different values of $a$.}
\end{figure}
We assume that the time-series $\{x_n\}$ is generated by an auto-regressive model of order 1
(AR(1)) (see e.g., \cite{Box-Jen}) 
\begin{equation}
x_{n+1}= a x_n + \xi_n,
\end{equation}
where $\xi_n$ are uncorrelated Gaussian random numbers with unit variance and $-1 < a < 1$ is a constant
which represents the coupling strength. The size and the sign of the coupling
strength sets whether successive values of $x_n$ are clustered or
spread, as illustrated in Fig.\ \ref{fig:data+}. 

In the case $a =0$ the process reduces to uncorrelated random numbers with
mean $\mu =0$ and variance $\sigma^2 =1$, whereas generally the process is
exponentially correlated $\langle x_n x_{n+k}\rangle = a^k <1 $ and has the
marginal PDF
 \begin{equation}
\rho(x_n) = \sqrt{\frac{1-a^2}{2\pi}}\exp\left(-\frac{1-a^2}{2}{x_n}^2 \right).
\end{equation}  
Since the size of the events is naturally measured in units of the standard deviation $\sigma(a)$ we introduce a new scaled variable $\eta =
\frac{d}{\sigma(a)} =d\sqrt{1-a^2}$.

 Applying the filter mechanism developed in Sec.
\ref{tfilter} we obtain the following expressions for the posterior PDF of extreme events and the posterior PDF of non-extreme events
\begin{eqnarray}
\rho(x_n|X(\eta),a)
&=&\frac{\sqrt{1-a^2}}{2\sqrt{2\pi}\rho^{\Theta}(a,\eta)}\exp\left(-\frac{1-a^2}{2}x_{n}^2\right)
\nonumber\\
&\it{}&\quad \mbox{erfc}\left(\frac{(1-a)x_{n}}{\sqrt{2}} +
\frac{\eta}{\sqrt{2}\sqrt{1-a^2}}\right), \label{marga}\\
\rho(x_n|\overline{X(\eta)},a)
&=&\frac{\sqrt{1-a^2}}{2\sqrt{2\pi}(1-\rho^{\Theta}(a,\eta))}\exp\left(-\frac{1-a^2}{2}x_{n}^2\right)
\nonumber\\
&\it{}& \left( 1 + \mbox{erf}\left(\frac{(1-a)x_{n}}{\sqrt{2}} +
\frac{\eta}{\sqrt{2}\sqrt{1-a^2}}\right)\right).\nonumber\\
\label{antima}
\end{eqnarray}
%
\subsection{Determining the precursor value\label{Det}}
Because of the Markov-property of the AR(1) model the probability for an event
at time $n+1$ depends only on the last value $x_n$, hence $k=1$ in Eq.\ (\ref{precursor}). Thus, we give an alarm for an extreme
event when an observed value $x_n$ is in an interval $ V_{pre}  =  [x_{pre} - \delta/2,x_{pre} + \delta/2] ; \label{I1}$
around the precursor value $x_{pre}$. We compute the precursor values $x_{I}$ and $x_{II}$ defined by Eq.\ (\ref{precursor}) according to the strategies
described in Sec. \ref{pre}. 

The maximum ${x_I}$ of \ma is given by the solution of the transcendental equation
\begin{eqnarray}
{x_I}(\eta) & = & \frac{\sqrt{2}}{\sqrt{\pi} (1+a)} \frac{\exp \left(-
\frac{1}{2}\left((1-a){x_I}  +\frac{\eta}{\sqrt{1-a^2}}\right)^{2}
\right)}{\mbox{erfc}\left({\frac{(1-a){x_I}}{\sqrt{2}} +
\frac{\eta}{\sqrt{2}\sqrt{1-a^2}}}\right)}. \nonumber \\ \label{trans}
\end{eqnarray}
Inserting the asymptotic expansion for large arguments of the complementary error function
\begin{eqnarray}
\erfc(z) &\sim &  \frac{\exp(-z^2)}{\sqrt{\pi}z} \left(1 + \sum_{m=1}^{\infty}
(-1)^m \frac{1 \cdot 3 ...(2m-1)}{(2z^2)^m }\right),\nonumber\\ 
&\it{}& \left( z \rightarrow \infty, |
\mbox{arg} z | < \frac{3\pi}{4} \right)\label{erfcapprox}
\end{eqnarray}
which can be found in \cite{Abram} we obtain:
\begin{eqnarray}
{x_{I}}(\eta) &\sim & \frac{-\eta}{2\sqrt{1-a^2} \left( 1 + 
\mathcal{O}
\left(\frac{1}{\eta ^2} \right)\right)}, \label{dhalbe}
\quad (\eta \rightarrow \infty).
\end{eqnarray}
%
\begin{figure}[t!!!]
\includegraphics[width=6cm, angle= -90]{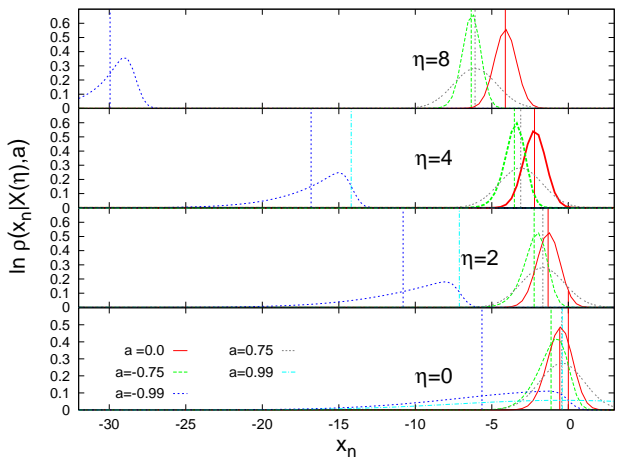} 
\caption[]{\small\label{fig:marginala-} (Color online)
The a posteriori PDFs for the AR(1) process for different values of $a<0$ and
$\eta$. The vertical lines represent the means. The PDFs become asymmetric for
$a \rightarrow -1$. (For $a=-0.99$ and $\eta \rightarrow \infty$ the marginal
PDFs becomes very flat and can hence not be distinguished from the x-axis in this figures).}
\end{figure}
Fig.\ \ref{fig:marginala-}  shows the posterior PDFs \ma according to
Eq.\ (\ref{marga}) for different values of $a$ and $\eta$. One can see
that the maximum of \ma moves  towards $- \infty$ with increasing size of
$\eta$ and $a \rightarrow 1$.
Although we can always formally define the maximum $x_I$ and the
mean~$\langle x_n \rangle$ as precursor values, one can argue that the
maximum of the distribution has no predictive power if  $a \rightarrow
1$. Since the variance of the posterior PDF increases immensely in this limit, the value of \ma in its
maximum does not considerably differ from the values in any other
point.

For large values of $\eta$ we can also assume that the maximum and the mean of \ma nearly coincide, i.e., 
\begin{equation}
 \langle x_n \rangle  \simeq {x_{I}} \sim  
\frac{-\eta}{2\sqrt{1-a^2} \left(1 + \mathcal{O} \left(\frac{1}{\eta^2} \right)\right)},\quad (\eta \rightarrow \infty), \label{dhalbe2}
\end{equation}
provided that \ma is not too asymmetric (i.e., $a$ is not close to $-1$). 
In the numerical tests in Sec.\ \ref{AR1_1} we will hence use the mean of the posterior
PDF as a precursor for strategy I, since it can be calculated explicitly by
evaluating the corresponding integral.

In order to determine $x_{II}$, the precursor for strategy II, we have to find
the maximum in $x_n$ of the likelihood 
\begin{eqnarray}
\rho(X(\eta)|x_n,a) 
& = &\frac{1}{2} \erfc\left(
\frac{(1-a)x_n}{\sqrt{2}} +
\frac{\eta}{\sqrt{2}\sqrt{1-a}}\right). \nonumber\\
\label{erfcII}
\end{eqnarray}
Since the complementary error function is a monotonously decreasing function of
$x_n$ we see that we do not have a well defined maximum~$x_{II}$, ( we will
thus denote $x_{II}:-\infty$) and that
the interval $V_-=\left( -\infty,x_- \right]$ with the upper limit $x_{-}$ represents the interval for raising alarms according to
strategy II.

\subsection{Testing the Performance of the Precursors \label{AR1_1}}

In order to test for the predictive power of the precursors specified above,
we  used two different methods to create
ROC-curves (see Sec.\ \ref{roc}). The first method consists in evaluating the integrals which lead to the rate
of correct and false predictions
\begin{eqnarray}
r_c(x_{pre},\eta,\delta) & = & \int_{V(\delta)} dx_{n}
\;\rho(x_{n}|X(\eta),a), \label{rcar} \\
r_f(x_{pre},\eta,\delta) & = & \int_{V(\delta)} dx_{n} \;\rho(x_{n}|\overline{X(\eta)},a). \label{rfar}
\end{eqnarray}
The second method consists in simply performing predictions on a time
series of $10^7$ AR(1) data and counting the number of extreme increments,
which could be predicted by using the precursors specified above.
For
different values of the correlation coefficients the data sets contained the
following numbers of extreme increments:
\begin{center}
\begin{tabular}{|c||c|c|c|c|}
\hline 
 & \multicolumn{4}{c|}{ number of increments of size}\\ \hline
$a$ & $\eta \geq 0$ & $\eta \geq 2$ & $\eta \geq 4$ & $\eta \geq 8$\\ \hline
\hline
-0.99 & 5000059 & 1579103 & 222858 & 310 \\ \hline
-0.75 & 5000563 & 1425146 & 162405 & 107 \\ \hline
0 & 5000417 & 786355 & 23370 & 0 \\ \hline
0.75 & 5000818 & 23377 & 0 & 0 \\ \hline
0.99 & 5001081 & 0 & 0 & 0 \\ \hline
\end{tabular}
\end{center}
In all cases, where the AR(1) correlated data sets contain increments, the empirically determined rates comply very well with the rates
obtained via the evaluation of Eqs.\ (\ref{rcar}) and (\ref{rfar}). For those
values of $a$ and $\eta$, which were not accessible for the numerical test, we evaluated the integrals in Eqs.\ (\ref{rcar}) and (\ref{rfar}).   

In the numerical tests for both strategies and also for the evaluation of the
integrals in Eqs.\ (\ref{rcar}) and (\ref{rfar}) according to strategy I, the
size of the precursory volume ranged from $ 10^{-6} $ to $ 4 $,
measured in size of the standard deviation of the marginal PDF of the AR(1)
process $\sigma (a) = 1/ \sqrt{1-a^2}$. As precursors according strategy I
we used the means of the a posteriori PDF. For the empirically created
ROC-plots according to strategy II we used the smallest values of the data
sets as precursors. 

The evaluation of the integrals in Eqs.\ (\ref{rcar}) and
(\ref{rfar}) was done in a slightly different way for strategy II. Since there
were no events in the data sets for certain value of $a$ and $d$ (as indicated
in the table above), one could argue that the data sets also did not
contain any precursor. From the previous section, we know that the theoretical precursor value according to
strategy II should be $x_{II} =- \infty$. Thus, we used a sufficiently small
value as a precursor and adjusted the size of the prediction interval in order
to capture all events. However, the resulting ROC-curves for strategy II
coincided with the curves obtained empirically, as far as they were available.
\begin{figure}[t!!!]
\includegraphics[width=6cm, angle= -90]{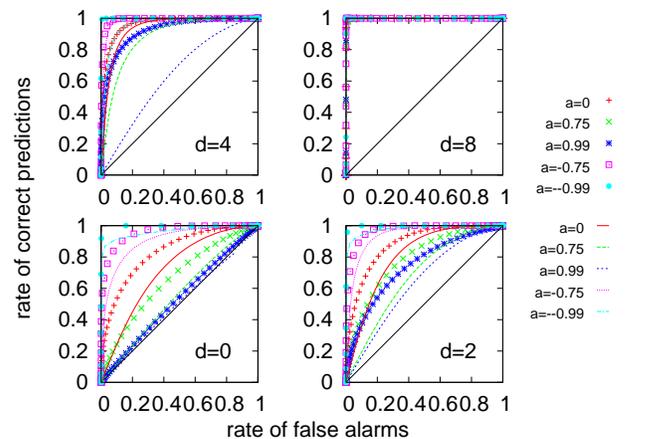} 
\caption[]{\small \label{fig:newstraroc} (Color online)
The  ROC-Curves made for the precursors of strategy I and II. The lines
represent the results of strategy I, the symbols correspond to predictions
made according to strategy II. In both cases the predictions were made within
$10^7$ AR(1)- correlated data. For the values of $\eta$ and $a$, where the
data sets contained no increments, we created the ROC-curves by
evaluating the integrals in Eqs.\ (\ref{rcar}) and (\ref{rfar}).}
\end{figure}

The resulting ROC-curves in Fig.\ \ref{fig:newstraroc} display the following
properties:

{\bf ad (Q1):}\hspace{3mm} The predictions according
to strategy II are better than the predictions according to strategy I for all
values of $a$ and $\eta$.

{\bf ad (Q2):}\hspace{3mm}
The ROC-curves display an increase of the quality of our
prediction with increasing size of the events $\eta$.

{\bf ad (Q3):}\hspace{3mm}
The ROC-curves in Fig.\ \ref{fig:newstraroc} show that the quality of the
predictions increases with decreasing correlation strength $a$.
Especially for $a=0$, when the predictions were made within completely
uncorrelated random numbers, the ROC curves are far better
than ROC curves for any random prediction.
 This is in agreement
with results reported 
in \cite{Sornette1} for
the prediction of signs of increments in uncorrelated random numbers, i. e.,
the case ($a=0, \eta =0)$. 

Intuitively, the result for {\bf (Q3)} can be understood easily by considering that increments are
not independent from the last observation. More precisely $x_{n+1} - x_n =
(a-1)x_n + \xi_n$, so that the known part of the increment $(a-1)x_n$ is the
larger, the smaller $a$.
In other words: if we consider a very small value of $x_n$ (small compared to
the mean) in an uncorrelated process, the probability that the next value will
be closer to the mean and hence lead to a large increment is high. Positive correlation hinders this effect,
since it causes successive values to be closer to each other. 

A formal explanation of the results {\bf (Q1)-(Q3)} is also given by an
asymptotic expression for the slope~$m(a,\eta, x_{pre})$ in the following section.
\subsection{Analytical discussion of the Precursor Performance
\label{analyse}}
In this section, we will try to understand the effects shown by the ROC-curves
in the previous section more detailed. Thus, we evaluate the asymptotic structure of the likelihood ratio as defined by Eq.\ (\ref{defm}) for different scenarios. 

In the case of the AR(1) process the slope of the ROC-curve in the vicinity of
the origin is given by 
\begin{eqnarray}
m(a,\eta, x_{pre}) \sim \quad
\frac{\left(1-\rho_{\Theta}(\eta)\right)}{\rho_{\Theta}(\eta)}\, r\bigl(x_{pre},\eta\bigr) ,\label{mAR}\\
\mbox{with}\quad r\bigl(x_{pre},\eta\bigr) =  \frac{\erfc\left(\frac{(1-a)x_{pre}}{\sqrt{2}} +
\frac{\eta}{\sqrt{2}\sqrt{1-a^2}}\right)}{1
+\mbox{erf}\left(\frac{(1-a)x_{pre}}{\sqrt{2}} +
\frac{\eta}{\sqrt{2}\sqrt{1-a^2}}\right)}.  \label{ratio}
 \end{eqnarray} 
{\bf ad (Q1):}\hspace{3mm}
We will first consider the behavior of the precursor according to strategy
II. As we saw in Sec. \ref{Det}, the optimal precursor value of strategy II is
the limiting case $x_{II}=-\infty$.

Since  
$\lim_{x_{pre} \rightarrow -\infty} r\bigl(x_{pre},\eta\bigr) =
\infty $ we find $\lim_{x_{pre} \rightarrow -\infty} m (a,\eta, x_{II}) = \infty$. Thus, we should expect  ROC-curves made with $x_{II} =-\infty$ to be tangent to the vertical axis of the curve and hence represent an ideal
predictability for all sizes of events and all possible correlation
strengths. However, for any finite precursor value of strategy I and strategy II we find non-ideal ROC-curves. 

\begin{figure}[t!!!]
\includegraphics[width=5.5cm, angle= -90]{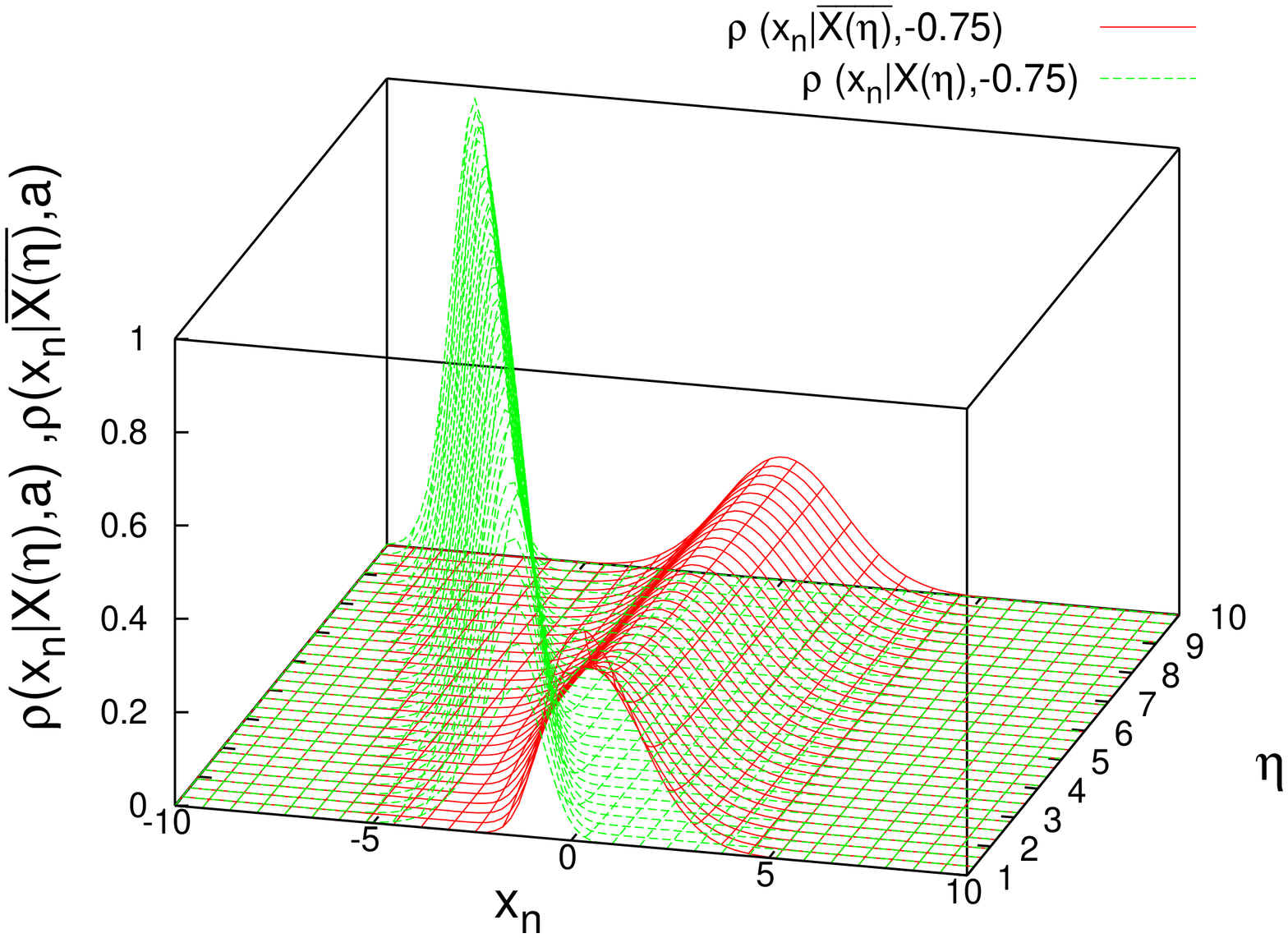} 
\caption[]{\small\label{fig:dreids} (Color online) \ma and \antima for $a
=-0.75$. The maximum of the posterior PDF to observe extreme events \ma which
   is used as precursor, moves towards $-\infty$ with increasing $\eta$
since $x_I \sim -\eta/(2\sqrt(1-a^2))$. Because the maximum of the failure
posterior PDF \antima remains at the origin, the values of
\antima which are observed at the precursor value ${x_{I}}$ decrease
according to the decrease of \antima as $x_n \rightarrow -\infty$.   
}
\end{figure}

Another way to understand the superiority of strategy II is to analyze the asymptotic behavior of the rate of correct predictions
\ma and the rate of false alarms,
\antima at the precursor value of
strategy I. For the following calculations we use an
approximation for the total probability to observe extreme events
%
\begin{eqnarray}
\rho_{\Theta} (\eta,a) & \sim &
\frac{\sqrt{1-a}}{\sqrt{\pi} }\;\frac{1}{\eta}\;
\exp\left(-\frac{\eta^2}{4(1-a)}  \right)\nonumber\\
&\it{}& 
\left( 1 +  \mathcal{O} \left(\frac{1}{\eta^2} \right)\right), \quad (\eta
\rightarrow \infty), \label{P}
\end{eqnarray}
which is derived in Appendix \ref{Ap1}.

Inserting the asymptotic expression  for $\rho_{\Theta}(\eta,a)$, the approximation of $x_I$ in
Eq.\ (\ref{dhalbe2}) and the asymptotic expansion of the complementary error function Eq.\ (\ref{erfcapprox})  into Eqs.\ (\ref{marga}) and
(\ref{antima}), we find the following expressions
\begin{eqnarray}
\rho \left(x_I|X(\eta),a \right) 
&\sim& \frac{\sqrt{1-a^2}\sqrt{1+a}}{\sqrt{\pi}} \frac{\left(1+\mathcal{O}\left(\frac{1}{\eta^2} \right)\right)}{\left(1+a +
\mathcal{O}\left(\frac{1}{\eta^2} \right)\right)},\nonumber\\
&\it{}&\quad  (\eta \rightarrow \infty).
 \label{qasym}  \\
\rho\left(x_I|\overline{X(\eta)},a,\right) & \sim &
\frac{\sqrt{1-a^2}}{\sqrt{2\pi}} \exp \left( -\frac{\eta^2}{8} \frac{1}{\left( 1 + 
\mathcal{O}\left(\frac{1}{\eta^2}
\right)\right)}\right) ,\nonumber\\
&\it{}& \quad
(\eta \rightarrow \infty)\quad .  \label{qbarapp}
\end{eqnarray}
Hence the value of \ma  at the precursor value approaches a constant for large
$\eta$, whereas the values of \antima decrease exponentially in this limit.
Fig.\ \ref{fig:dreids} illustrates this effect for
the case $a=-0.75$. The maximum of the failure PDF
remains at the origin for $\eta \rightarrow \infty$. Thus the
values of this PDF which are observed at the decreasing
precursor value $x_{I}\propto \frac{-\eta}{2\sqrt{1-a^2}}$ decrease
according to the shape of the distribution. This explains also the success of strategy II. Since the precursor value obtained by
strategy II is the smallest possible value, strategy II seems to focus on the
minimization of the failure rate. Note that by "minimization of
the failure rate", we understand here a minimization of the integrand in Eq.
(\ref{rfar}), while the alarm interval of size $\delta$ remains constant.
The fact that in this point the corresponding value of \ma is also far away
from the maximum of \ma does apparently not influence the outcome of the
prediction. 


{\bf ad (Q2):}\hspace{3mm} In the following calculation we will obtain the
asymptotic form of the likelihood ratio for large events. 
 Inserting the asymptotic form of the probability $\rho_{\Theta}(\eta,a)$ provided by
Eq.\ (\ref{P}),
and using the asymptotic expansion of
the complementary error function in Eq.\ (\ref{erfcapprox}), the likelihood ratio reads 
\begin{widetext}
\begin{eqnarray}
m(a,\eta, x_{pre}) & \sim & \frac{1}{2\sqrt{1-a}}\;\;  \frac{\eta\exp
\left(\frac{\eta^2}{4(1-a)} -z(\eta,a)^2 \right) \left(1+\mathcal{O} \left(\frac{1}{\eta^2} \right)
\right)} {z(\eta,a)\left(1 +
\mathcal{O}\left(\frac{1}{\eta^2} \right)\right) +
\mathcal{O}\left(\exp(-z(\eta,a)^2)\right)}
+\mathcal{O}\left(\frac{\exp(-z(\eta,a)^2)}{z} \right),\nonumber\\
&\it{} & \quad \quad (\eta \rightarrow \infty), \;(z(\eta,a)
\rightarrow \infty)
\quad \quad \mbox{with} \quad z(\eta,a)  = \frac{(1-a)}{\sqrt{2}}x_{pre} + \frac{\eta}{\sqrt{2}\sqrt{1-a^2}}.\label{m3}
\end{eqnarray} 
\end{widetext}
Note that the limit $z(\eta,a)\rightarrow \infty$ corresponds to the
limit $\eta \rightarrow \infty$ in the context of (Q2), but we can
also interpret it as the limit $a \rightarrow \pm 1$ in the context of (Q3) if
$\eta\neq0$.

The expression in Eq.\ (\ref{m3}) tends to infinity in the limit $\eta
\rightarrow \infty$, if the argument of the exponential function in Eq.\ (\ref{m3})
{\small 
\begin{eqnarray}
f(x_{pre},a,\eta) & = & \frac{\eta^2}{4(1-a)} -\left( \frac{(1-a)x_{pre}}{\sqrt{2}} +
\frac{\eta}{\sqrt{2}\sqrt{1-a^2}}\right)^2, \nonumber\\\label{arg}
\end{eqnarray}
}
is positive. This is indeed the case for every precursor value $x_{pre}<0$. Therefore, for both strategies of prediction, the slope $m(x_{pre},a,\eta)$ increases as
a squared exponential with increasing size of the events $\eta$ according to
Eq.\ (\ref{m3}). Hence, the considerations of
Sec. \ref{roc} hold for our example, according to which an event is the better predictable
the more rare it is.

{\bf ad (Q3):}\hspace{3mm} One can also calculate the asymptotic behavior of
the likelihood ratio for $a \rightarrow \pm 1$. The limit
$z(\eta,a) \rightarrow \infty$, which is relevant for the asymptotic
form in  Eq.\ (\ref{m3}), can also be interpreted as the limit $a \rightarrow
\pm$ 1. We assume that  $\eta$ is big enough, e.g., $\eta >2$,
such that Eq.\ (\ref{P}), which enters into Eq.\ (\ref{m3}), is a useful approximation. One can now discuss again the argument of the exponential function in Eq.\ (\ref{arg}).

\begin{figure}[t!!!]
\includegraphics[width=6cm, angle= -90]{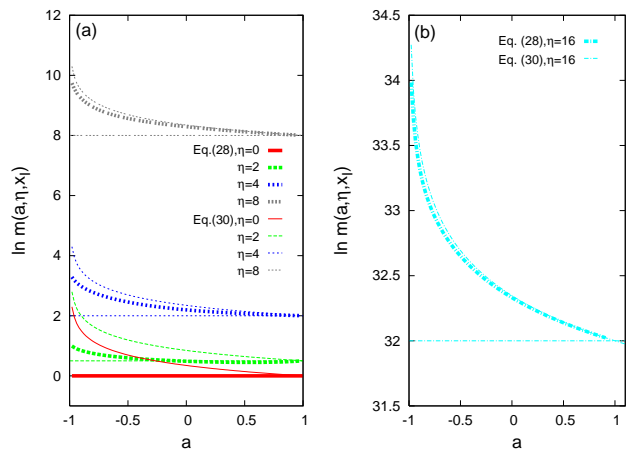} 
\caption[]{\small\label{fig:adepm} (Color online)  The bold lines show the
dependence of the slope $m(x_I,a,\eta)$ on the coupling strength according to
Eq.\ (\ref{m3}). The thin lines display  the asymptotic behavior, given by
Eq.\ (\ref{straIa}). The constant lines represent the values, obtained from
Eq.\ (\ref{straIa}) in the limit $a\rightarrow 1$. Fig.\ (b) illustrates, that
this asymptotic expression becomes better in the limit $\eta\rightarrow
\infty$, since in this limit the higher order terms in the approximation vanish even faster.
}
\end{figure}

Inserting the precursor of strategy I (as given by Eq. ref{}), one obtains
$f(x_{I},a,\eta)  =   \frac{\eta^2}{8}$, hence
\begin{eqnarray}
 m(a,\eta, x_{I})&\rightarrow &  \sqrt{\frac{2}{1+a}}\exp
\left(\frac{\eta^2}{8}\right), 
 \quad  (z(\eta,a)
\rightarrow \infty).\nonumber\\ \label{straIa}
\end{eqnarray}
As $a\rightarrow 1$, this expression converges to  $\exp\left(\eta^2/8\right)$. As $a\rightarrow -1$, this expression
approaches infinity as $m(1,\eta, x_{I}) \sim 1/\sqrt{1+a}$.
Fig. \ref{fig:adepm}(a) illustrates this behavior.
 Fig.\ \ref{fig:adepm}(b) shows that the asymptotic expression in Eq.\ (\ref{straIa}) becomes
better in the limit $\eta\rightarrow \infty$, since in this limit the higher
order terms of the approximation vanish even faster.

For the theoretical precursor of strategy II $x_{II} = -\infty$ the slope would be independent of the value of the coupling strength if
the exact precursor of strategy II could be used. For any real precursor value
of strategy II $x_{II} = \mbox{const.} <0$,
Eq. (\ref{arg}) reads
\begin{eqnarray}
f(x_{II},a,\eta) & \sim & \frac{\eta^2}{2(1-a)} \left( \frac{1}{2} - \frac{1}{1+a}
\right)\nonumber\\
&\it{}&  + \mathcal{O}\left((1-a)
\right), \quad(a \rightarrow 1).
\end{eqnarray}
This expression approaches a small negative value close to zero in the point $a
= 1$. Hence, we find $m(a,\eta,x_{II})  \sim 1$, as $a\rightarrow 1$.

In the limit $a \rightarrow -1$ and for any finite precursor value $x_{II}=\mbox{const.}<0$ Eq.\ (\ref{arg}) reads
\begin{eqnarray}
f(x_{II},a,\eta)  &\sim& \frac{\eta^2}{4} \left(\frac{1}{2} -
\frac{1}{1-a^2} \right) -\frac{2x_{II}\eta}{\sqrt{1-a^2}} -2x_{II}^2\nonumber\\
&\sim& -\frac{1}{1-a^2}\frac{\eta^2}{4} -\frac{2x_{II}\eta}{\sqrt{1-a^2}} -2x_{II}^2.
\end{eqnarray}
If the precursor is sufficiently small, e.g $x_{II} <-\eta/(4\sqrt{1-a^2})$,
this expression is positive and hence $m(a,\eta,x_{II}) \rightarrow
\infty$, as $a \rightarrow -1$.
\begin{figure}[t!!!]
\includegraphics[width=6cm]{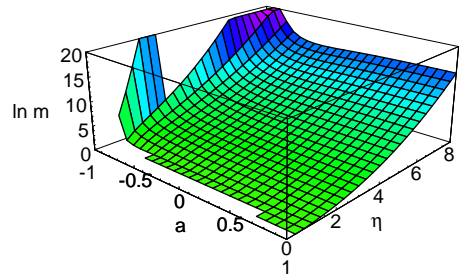} 
\caption[]{\small\label{fig:addepende} (Color online) The asymptotic dependence of the slope
$m(x_I,a,\eta)$ on the coupling strength and the event size, if the
precursor of strategy I is used.}
\end{figure}
Hence, the asymptotic expressions of the likelihood ratio are able to describe the
behavior of the ROC-curves, shown in the previous
section. Fig.\ \ref{fig:addepende} combines the dependence of the likelihood ratio
on the event size and the correlation strength. One can see that the
influence of the event size on the likelihood ratio is dominating, as long as
one does not approach the singularity at $a \rightarrow -1$.
\section{Application: wind speed measurements}\label{wind}

As an illustration of the preceeding considerations and also in order
to demonstrate the usefulness of the benchmarks derived for AR(1)
processes, we study here time series data of wind speed measurements. The
data are recorded at 30m above ground by a cup anemometer with a
sampling rate of 8 Hz in the Lammefjord site of the Ris\o\ research
center \cite{winddata}. Wind speed data are evidently non-stationary
and strongly correlated, so
that, e.g the principle of persistence yields surprisingly accurate
forecasts: the very simple prediction scheme $\hat{x}_{n+1}=x_n$ is
almost as accurate as an AR(20) model fitted on moving windows (in
order to take non-stationarity into account) or order-10 Markov
chains\cite{Euromech}. The amplitude of the fluctuations around a time
local mean value are proportional to this mean value, i.e., there is
statistical evidence that the noise in this process is
multiplicative. However, when subtracting the time local mean (more
precisely, performing a high-pass filtering with a Gaussian kernel 
with a standard deviation of 75 time steps), we receive data for which
it is reasonable to fit an AR(1) process. When doing so, we find a
coefficient $a\approx 0.94$. 

Turbulent gusts, i.e., sudden increases of the
wind speed, are relevant events, e.g for the save operation of wind
turbines, for aircrafts during take-off and landing, and for all
wind-driven sports activities. In previous work\cite{Physa} we were
therefore concerned with their prediction, where we were studying the
performance of a Markov chain model. Here, we will restrict ourselves
to the simpler (and less appropriate) AR(1)-philosophy: The current
state of the process generating the wind time series is assumed to be
fully specified by the last observation $x_n$, and the event is
assumed to be characterized by the upward jump of the wind speed in a
single time step by more than $g$ m/s. 

\subsection{Determining the precursor value}
If we extract from the data set
all subsequences of data where such a jump is present, then we can, in
principle, construct empirically the distribution $p(x_n|g)$, which
corresponds to $\rho({\mathbf x}_{(n,k)}|X)$ of strategy I. 
\begin{figure}[h!!!]
\centerline{\includegraphics[width=8cm]{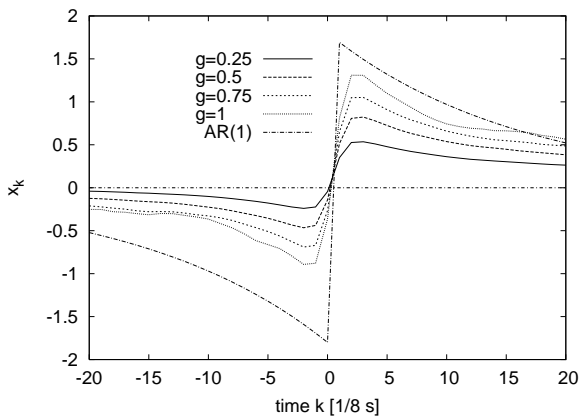}}
\caption{\label{fig:gustprofiles} 
The profiles obtained from the mean of $p(x_{n+k}|g)$ for gust events of
amplitude $g$. Also shown is the theoretical 
profile for an AR(1) process with $a=0.94$}
\end{figure}
\begin{figure}[h!!!]
   \centerline{\includegraphics[width=8cm]{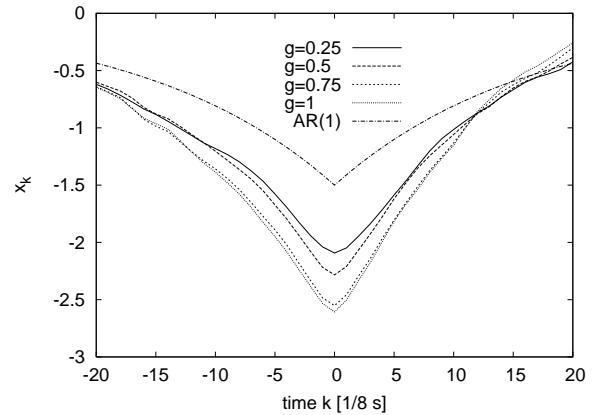}}
\caption{\label{fig:precursors2}
The profiles obtained from the maxima of $p(g|x_{n+k})$ for gust events of
amplitude $g$. Also shown is the theoretical profile for an AR(1) process
with $a=0.94$.}
\end{figure}
In Fig.\ \ref{fig:gustprofiles} we show instead the mean value of
$p(x_{n+k}|g)$ for $k=-20,\ldots,20$, i.e., we show the mean profile
of gusts of strength $g$. Otherwise said, this is an average of all
those time series segments, which (in shifted time) fulfill
$x_1-x_0>g$, so that the part of these segments with $k\le 0$ is what
one would call naively a precursor of a gust event. This has to be
compared to the values $x_{n+k}$ which we find when we focus on the
maximum $x_{II}$ in $x_n$ of $p(g|x_n)$ which corresponds to the conditional
probability $\rho(X|x_n)$ of strategy II.
More specifically, in Fig.\ \ref{fig:precursors2} we show the profiles $\langle
x_{n+k}\rangle|_{x_n=x_{II}}$ , where $x_{II}$ is defined by
$p(g|x_{II})=max_{x_{n}}$. In even different words, the value plotted at $k=0$ is the
value $x_n$ for which $p(g|x_n)$ is maximal, and at the preceeding and succeeding
time steps we show the average over all time series segments which fulfill
$x_n=x_{II}$ is some precision. These profiles differ from the precursors shown
before, as we have to expect for an AR(1)-model: In a perfect AR(1) process,
the precursors equivalent to those in Fig.\ \ref{fig:gustprofiles} would show a
jump larger than
$g$ from $k=0$ to $k=1$, with $x_0=-x_1$, and with $x_k=a^k x_0$ for
$k<0$, and $x_k=a^k x_1$ for $k>1$. For the same idealized process, one
expects Fig.\ \ref{fig:precursors2} to show curves given by
$x_k=a^{|k|} x_{II}$ for all $k$. Evidently, the wind data show a qualitatively very
similar behavior, whereas, however, additional correlations are visible. 

\subsection{Testing for predictive power}
\begin{figure}[h!!!]
\centerline{\includegraphics[width=6cm, angle =-90]{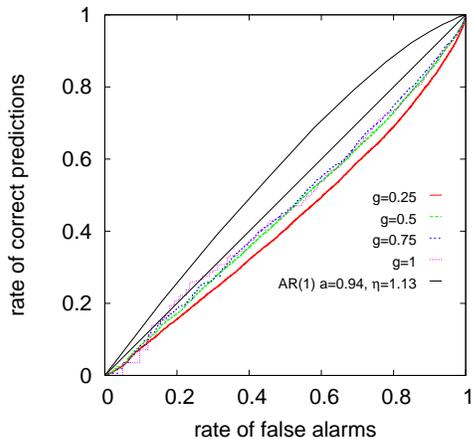}}
\caption{\label{fig:ROCwind1} (Color online)
The ROC curves using strategy I, exploiting $p(x_n|X)$ and maximizing the hit
rate. Evidently, the rate of false alarms exceeds the hit rate.}
\end{figure}

\begin{figure}[t!!!]
\centerline{\includegraphics[width=6cm, angle =-90]{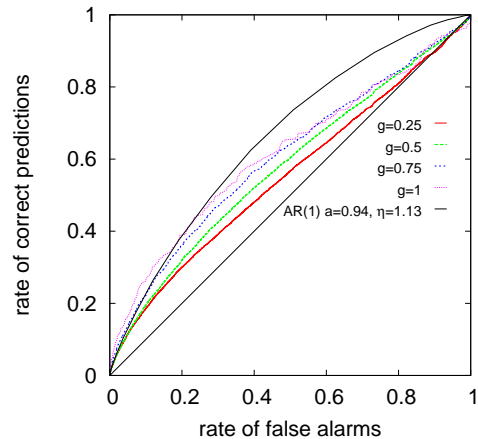}}
\caption{\label{fig:ROCwind2} (Color online)
The ROC-curves for the prediction of jumps of amplitude larger than $g$
for the wind data. Strategy II exploits $p(X|x_n)$ which minimizes 
the false alarm rate and
performs the better the larger $g$.
}
\end{figure}

The ROC-curves for the two prediction strategies are shown in
Fig. \ \ref{fig:ROCwind1} and \ref{fig:ROCwind2}. As expected, the minimization
of false alarms (strategy II) is here superior, as strategy I has no predictive
power. The latter is consistent with the observed value $a\approx 0.94$ and
the results for the AR(1) process.

In order to compute the ROC-curves we use 
the following numerically expensive but 
theoretically best justified algorithm:
In theory, we want to generate an alarm if the current observation
$x_n$ lies in an interval $V$ which is defined by the
subset of the ${\mathbb R}$ where either $p(g|x_n)$ or $p(x_n|g)$ exceeds
some threshold $0\le p_c \le 1$. We assume that both 
conditional PDFs are smooth in $x_n$. 

We can locally approximate
$p(g|x_n)$ by searching all similar states 
${x_j}$, with $|x_n-x_j|<\epsilon$ and counting the relative number 
of events in this set of states. When this number exceeds $p_c$, 
we give the alarm and can see whether it is a hit or a false alarm.

In order to evaluate $p(x_n|g)$ we first create the set of all states $x_{e}$
which  are preceeding an event, and then compute the fraction of
these which is $\epsilon$-close to the current state $x_n$. Since
this fraction evidently depends on the value of $\epsilon$, we should
introduce a normalization. However, in order to create the ROC statistics we just have to introduce a threshold which runs from 0 to
the largest value thus found. Both schemes can be straightforwardly
generalized to situations where the current state of the process is
defined by a sequence ${\mathbf x}_{(n,k)}$ of $k$ past measurements
$(x_{n-k+1}, x_{n-k+2}, \ldots,,x_{n-1},x_n)$, e.g., for an AR(2) model
$k=2$,  whereas in \cite{Physa} we were using $k=10$ for a Markov chain of
order 10. 

Since the wind speed data are strongly correlated, $a\approx 0.94$, it is not
possible to predict the increments of the data sufficiently well.  This
corresponds to the previously derived results for the AR(1) model in the
limit $a \rightarrow 1$. However, we also find deviations from the theoretical
ROC-curve for $a=0.94$, which is additionally plotted in Figs.\ \ref{fig:ROCwind1}
and \ref{fig:ROCwind2}. These deviations show that the AR(1) model is not
able to describe the wind data completely.

The wind data also show the increase of predictability with increasing event
size. This suggests that this effect is more general and not
limited to the class of AR(1) models. Again, we also observe that strategy II
is superior to strategy I.
\section{Extreme Increments in long-range correlated Processes \label{long}} 
We studied the same questions which are described before, in
long-range correlated processes.  Since the precursors we were interested in
live on a very short time scale (one step before the event), one should not
expect long-range correlations to lead to qualitatively different results for
the aspects we were interested in. The results obtained in this section
support this assumption.

There are various definitions of long-range
correlation. Typically long-range correlation in a time series is characterized by the exponent $0<\gamma_c<1$ of
the power-law decay of the autocorrelation function as a function of the time
t
\begin{eqnarray}
C_x(t) & = & \langle x_n x_{n+t}\rangle = \frac{1}{N-t} \sum_{n=1}^{N-t} x_n
x_{n+t} \sim t^{-\gamma_c}
\end{eqnarray}
The correlation coefficient $\gamma_c$ is controlling, how fast the correlations decay.

We study the predictability of increments numerically by applying the
prediction strategies described in Sec.\ \ref{pre}. 
The data used for this numerical study were generated as described in
\cite{Eduardosref} and used in\cite
{Eduardo1}: Imposing a power-law decay on the Fourier spectrum,
\begin{eqnarray}
f_x(k) \propto k^{-\beta}
\end{eqnarray}
with $0<\beta<0.5$ and choosing phase angles at random one obtains through an
inverse Fourier transform the long-range correlated time series in $x$ with
$\gamma_c=1-2\beta$. The data are Gaussian distributed with $ \langle x \rangle
=0$, $\sigma=1$. Having specified the power spectrum or, correspondingly, the
autocorrelation function for sequences of Gaussian random numbers means to
have fixed all parameters of a linear stochastic process. Hence, in principle
the coefficients of an autoregressive or moving average process can be
uniquely determined, where, due to the power-law nature of the spectrum and
autocorrelation function the order of either of these models have to be
infinite \cite{Brockwell,Box-Jen}. 
Thus, the effects which we observed for this
ARMA($\infty$, $\infty$) model should be valid for the whole class of linear long-term
correlated processes.
\begin{figure}
\includegraphics[width=6cm, angle= -90]{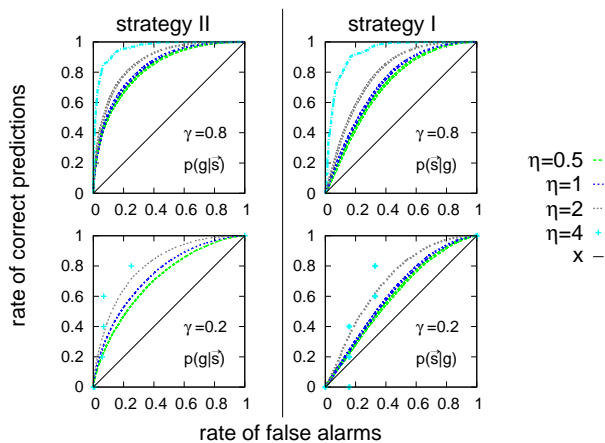} 
\caption[]{\small\label{fig:longrange} (Color online)
ROC-curves for the ARMA($\infty$,$ \infty$) processes  with
$\gamma_c=0.2$ and$\gamma_c=0.8$.}
\end{figure}
The ROC-curves in Fig.\ \ref{fig:longrange}, which are generated from the
long-range correlated data are very similar to the ones for the AR(1)
process in terms of the question we want to study. 

{\bf ad (Q1):}\hspace{3mm}
The ROC-curves obtained by
using strategy II are superior to the curves resulting from strategy I.

{\bf ad (Q2) and (Q3):}\hspace{3mm}
 The quality of the prediction also
increases with increasing event size and decreasing correlation. 

Hence we observe the same effects which we described before for the AR(1)
process and the wind speed data in a long range correlated ARMA($\infty$, $\infty$) process.

\section{Conclusions\label{con}}
We studied the predictability of extreme increments in an AR(1) correlated
process, in wind speed data and in a long-range correlated ARMA process. 
To measure the quality of the prediction we used the ROC-curve and additionally
the slope of the ROC-curve in the vicinity of the
origin as a summary index. This so called {\sl likelihood ratio},
characterizes particularly the behavior in the limit of low false-alarm rates. 

In the case of the AR(1) process we could construct the posterior PDF and the
likelihood analytically
from a given joint PDF and hence we were able to obtain the asymptotic
behavior of the likelihood ratio analytically. In the case of the two other
examples, we constructed the posterior PDFs numerically. The resulting distributions were then
used to determine precursors according to two different strategies of prediction. 

In all examples we studied the aspects : {\bf (Q1)} Which is
the best strategy to choose precursors? {\bf (Q2)} How does the
predictability depend on the event size? {\bf (Q3)} And how does the
predictability depend on the correlation? The results can be summarized as follows:

{\bf ad (Q1):}\hspace{3mm} 
Strategy I, the a posteriori approach,  maximizes  the rate of
correct predictions, while strategy II focuses on the
minimization of the rate of false alarms.
For the example of 
the AR(1) process one can show that strategy II is the optimal strategy to make predictions. 
For other stochastic processes, it is
not in general clear which of the two strategies leads to a better
predictability. However, the application to the prediction of wind speeds and
the numerical study within long-range correlated data reveals that also for
these examples better results are obtained by predicting according to strategy
II.

{\bf ad (Q2):}\hspace{3mm} 
For all examples studied, we
observe an increase of predictability with increasing size of the events. This
phenomenon which is also reported in the literature
\cite{Physa,Goeber,Johnson1}, can be better studied by investigating the asymptotic behavior of our summary index. In the case of the AR(1) process we showed explicitly that the likelihood ratio
increases as a squared exponential with increasing event size. 
In Sec.\ \ref{roc} we discussed for a general stochastic process that this
effect appears,  if the PDFs of the studied process fulfill certain conditions.

{\bf ad (Q3):}\hspace{3mm} 
For the AR(1) process and the long-range correlated data we observe that the correlation of
the data is  inversely proportional to the quality of the predictions. The
ROC-curves for the wind data, which we assume to be a strongly correlated AR(1)
process with correlation strength $a=0.94$, display also a bad
predictability. This effect is  due to the special definition of the events as
increments. The asymptotic expression for the likelihood ratio  in Eq.\
(\ref{m3}) provides us also with a formally understanding of the $a$-dependence. 

All the considerations made in this contribution are made for a very simple
but general method. In order to make predictions, we use the largest maximum of
the a posterior PDF or the likelihood. For multimodal distributions, one can
think about more sophisticated methods, which take into account also other
maxima of the distribution. Furthermore, we investigate only stationary
processes in these contributions. It remains to be studied, whether the answers, obtained to the questions
({\bf Q1})-({\bf Q3}) are also valid for non-stationary processes or multimodal
distributions. \\
\begin{acknowledgements}
E. G. A. was supported by CAPES (Brazil).
\end{acknowledgements}
\begin{appendix}
\section{Obtaining an asymtotic form of the total probability to find
increments of size $\eta$ \label{Ap1}}
The total probability $\rho_{\Theta}(\eta,a)$ to find increments of size $\eta$ can be obtained by
integrating the pre-form of the posterior probability in Eq.\ \ref{int1}. For the
example of the AR(1) process the corresponding integral reads
\begin{eqnarray}
\rho_{\Theta}(\eta,a)&=&\int_{-\infty}^{\infty}\frac{\sqrt{1-a^2}}{2\sqrt{2\pi}}\exp\left(-\frac{1-a^2}{2}x_{n}^2\right)
\nonumber\\
&\it{}&\quad \mbox{erfc}\left(\frac{(1-a)x_{n}}{\sqrt{2}} +
\frac{\eta}{\sqrt{2}\sqrt{1-a^2}}\right).
\end{eqnarray}  
In the special case $\eta=0$ one can find the analytical form of the total
probability 
$\rho_{\Theta}(0,a)$ using again an integral identity from \cite{Pru}. The resulting value $ \rho_{\Theta}(0,a) = 1/2$ corresponds to the
intuitive expectation one would have, since for $\eta=0$ the condition of our
extreme event is always fulfilled if $x_{n+1}$ is larger than
$x_{n}$. This special case of predicting the sign of increments in
uncorrelated data is discussed in \cite{Sornette1}.

For $\eta \neq 0$, we can find an asymptotic form of the total probability $
\rho_{\Theta}(\eta,a)$ via
evaluating the mean of the posterior PDF.
An analytic expression of the mean can be obtained using an integral representation from \cite{Pru}
\begin{equation}
\langle x_{n}\rangle = \frac{-\exp \left(-\frac{\eta^2}{4(1-a)}  \right)}{2 \sqrt{\pi}
\sqrt{1+a}\; \rho_{\Theta}(\eta, a)}, \label{anamean_ar}
\end{equation}
%
For large values of $\eta$ we can also assume that the maximum and the mean of \ma nearly coincide, i.e., 
\begin{equation}
 \langle x_n \rangle  \simeq {x_{I}} \sim  
\frac{-\eta}{2\sqrt{1-a^2} \left(1 + \mathcal{O} \left(\frac{1}{\eta^2} \right)\right)},\quad (\eta \rightarrow \infty), \label{dhalbe2}
\end{equation}
provided that \ma is not too asymmetric (i.e., $a$ is not close to
$-1$). Using this approximation, we find the following asymptotic form of the
total probability to find increments of size $\eta$

\begin{eqnarray}
\rho_{\Theta} (\eta,a) & \sim &
\frac{\sqrt{1-a}}{\sqrt{\pi} }\;\frac{1}{\eta}\;
\exp\left(-\frac{\eta^2}{4(1-a)}  \right)\nonumber\\
&\it{}& 
\left( 1 +  \mathcal{O} \left(\frac{1}{\eta^2} \right)\right), \quad (\eta
\rightarrow \infty). \label{P}
\end{eqnarray}

\section{Transformation of extreme increments into extreme values}
\label{trafo}
We show how to relate the 
results obtained using the definition of extreme events as extreme
increments~($x_{n+1}-x_n \geq d$, as in Eq.~(\ref{e0})) to the case when
extreme events are defined as extreme values ($y_{n+1} \geq d$) which exceed a
certain threshold $d$, for ARMA(p,q) processes.  An ARMA(p,q) model is defined
as~\cite{Box-Jen} 
\begin{equation}\label{eq.arma}
\Phi(B) x_n = \theta(B) \xi_n,
\end{equation}
where~$\{\xi\}$ correspond to white noise and
\begin{eqnarray*}
\Phi(B) & = &  1 - \Phi_1 B - \Phi_2 B^2 - ... - \Phi_p
B^p,\\
\theta(B) & = &  1 + \theta_1 B + \theta_2 B^2 + ... +
\theta_q B^q, 
\end{eqnarray*}
with $B^j x_n =  x_{n-j}$. Searching for extreme increments in a time
series~$\{x\}$ is equivalent to search for extreme values in the time
series~$\{y\}$, defined through the transformation 
\begin{equation}\label{eq.transf}
y_{n+1}  =  x_{n+1} - x_n.
\end{equation}

Assuming that $\{x\}$ is described by an ARMA(p,q) process defined by
Eq.~(\ref{eq.arma}), and inserting Eq.~(\ref{eq.transf}) in
Eq.~(\ref{eq.arma}), one obtains that $\{y\}$ is described by an
ARMA(p,q+1) model with 
the following transformed coefficients
\begin{eqnarray}\label{eq.identification}
\Phi^\dagger_i & = &\Phi_i \quad i=1,2,...p\quad, \nonumber\\
\theta^\dagger_i &=& \theta_i - \theta_{i-1} \quad i=1,2,...q \quad,\nonumber\\
 \theta_{q+1}^\dagger & = & \theta_q \quad.
\end{eqnarray}
 Due to the transformation~(\ref{eq.transf}) the precursory structure
 equivalent to the one used in Sec.~\ref{AR1} is obtained choosing\footnote{We
 assume $x_0=0$,  which is the mean value of $\{x\}$. This assumption is
 irrelevant for large values of~$n$.}  
\begin{equation}\label{eq.pre2}
y_{pre}=\sum_{j=0}^{n} y_j - x_0= x_n.
\end{equation}
With this choice of precursory structure and the corresponding
transformation of the process (Eq.\ (\ref{eq.transf})), the results obtained for
extreme increments can be transfered to the case of extreme values. In
particular, for the case of AR(1) processes (which corresponds to an ARMA(1,0))
discussed in Sec.~\ref{AR1}, all results are also valid for an ARMA(1,1)
process with the precursor given by~(\ref{eq.pre2}) and events defined
as extreme values. E.g the alarm strategies consist in this case in raising
an alarm whenever $y_{pre}$ falls near the precursor values given in
Eq.~(\ref{precursor}). 
\end{appendix}

\end{document}